\documentclass[conference]{IEEEtran}

\IEEEoverridecommandlockouts
\usepackage{booktabs}
 \usepackage[table,xcdraw]{xcolor}
\usepackage[hidelinks]{hyperref}
\usepackage{wrapfig}
\usepackage{verbatim} 
\usepackage{amsmath,amssymb,amsthm} 
\usepackage{listings} 
\usepackage{color}
\usepackage{graphicx}
\usepackage{enumerate}
\usepackage{paralist}

\usepackage[oldcommands,linesnumbered,boxed]{algorithm2e}
\setalcapskip{1em} 
\newcommand{\martinAlgFontsize}{\fontsize{8.0}{8.0}\selectfont}
\SetAlFnt{\martinAlgFontsize}
\newcommand{\martinAlgCapFontsize}{\fontsize{9.0}{9.0}\selectfont}
\SetAlCapNameFnt{\martinAlgCapFontsize}
\SetAlCapFnt{\martinAlgCapFontsize}

\newcommand{\martinListingFontsize}{\fontsize{8}{8}\selectfont}

\newcommand{\ackContent}{
    This work has been supported by the European Union's Horizon 2020 research and innovation programme under grant No 739551 (KIOS CoE). 
    To appear in Proceedings of the MaxSAT Evaluation 2019 (MSE'19),  \url{https://maxsat-evaluations.github.io/2019/}.
}

\newcommand{\cost}{\varphi} 
 
\newcommand{\form}{f_{G}}

\newcommand{\tool}{META4ICS } 
\newcommand{\toolx}{META4ICS}

\begin{document}

\thispagestyle{plain}
\pagestyle{plain}

\title{MaxSAT Evaluation 2019 - Benchmark: \\Identifying Security-Critical Cyber-Physical Components in Weighted AND/OR Graphs}

\author{   
	\thanks{\ackContent}
	\IEEEauthorblockN{
		Mart\'in Barr\`ere\IEEEauthorrefmark{1},         
		Chris Hankin\IEEEauthorrefmark{1}, 
		Nicolas Nicolaou\IEEEauthorrefmark{2}, 
		Demetrios G. Eliades\IEEEauthorrefmark{2}, 
		Thomas Parisini\IEEEauthorrefmark{3}		
	}
	\IEEEauthorblockA{\IEEEauthorrefmark{1}Institute for Security Science and Technology, Imperial College London, UK
		\\\{m.barrere, c.hankin\}@imperial.ac.uk}
	\IEEEauthorblockA{\IEEEauthorrefmark{2}KIOS Research and Innovation Centre of Excellence, University of Cyprus
		\\\{nicolasn, eldemet\}@ucy.ac.cy}		
	\IEEEauthorblockA{\IEEEauthorrefmark{3}Department of Electrical and Electronic Engineering, Imperial College London, UK
		\\\{t.parisini\}@imperial.ac.uk}		
}

\maketitle

\begin{abstract}
This paper presents a MaxSAT benchmark focused on identifying critical nodes in AND/OR graphs. We use AND/OR graphs to model Industrial Control Systems (ICS) as they are able to semantically grasp intricate logical interdependencies among ICS components. However, identifying critical nodes in AND/OR graphs is an NP-complete problem. We address this problem by efficiently transforming the input AND/OR graph-based model into a weighted logical formula that is then used to build and solve a Weighted Partial MaxSAT problem. The benchmark includes 80 cases with AND/OR graphs of different size and composition as well as the optimal cost and solution for each case. 
\end{abstract}

\newcommand{\content}{sections} 

\section{Problem overview}
\label{sec:intro}

Over the last years, Industrial Control Systems (ICS) such as water treatment plants and energy facilities have become increasingly exposed to a wide range of cyber-physical threats, having massive destructive consequences. Our work is focused on security metrics and techniques that can be used to measure and improve the security posture of ICS environments~\cite{Barrere2019}. We use AND/OR graphs to model these systems as they allow more realistic representations of the complex interdependencies among cyber-physical components that are normally involved in real-world settings \cite{Desmedt2002}, \cite{Desmedt2004}. In that context, we have designed a security metric, detailed in~\cite{Barrere2019}, whose objective is to identify the set of critical AND/OR nodes (ICS network components) that must be compromised in order to disrupt the operation of the system, with minimal cost for the attacker. 

From a graph-theoretical perspective, our security metric looks for a minimal weighted vertex cut in AND/OR graphs. This is an NP-complete problem as shown in \cite{Desmedt2004, DesmedtUsingFF, Souza2013}. While well-known algorithms such as Max-flow Min-cut \cite{Ford1962} and variants of it could be used to estimate such metric over OR graphs in polynomial time, their use for general AND/OR graphs is not evident nor trivial as they may fail to capture the underlying logical semantics of the graph. In that context, we take advantage of state-of-the-art MaxSAT techniques to address our problem.

\section{Simple example}

\begin{figure}[!h]
	\centering
	\includegraphics[scale=0.28]{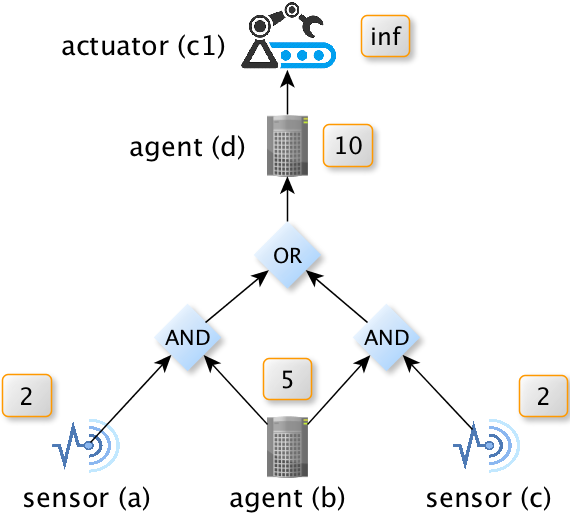}	
	\caption{AND/OR graph with sensors, software agents and actuators}
	\label{fig:simple-example1}
\end{figure}

Let us consider a simple ICS network whose operational dependencies are represented by the AND/OR graph shown in Figure \ref{fig:simple-example1}. 
The graph reads as follows: the actuator $c1$ depends on the output of software agent $d$. Agent $d$ in turn has two alternatives to work properly; it can use either the readings of sensor $a$ and the output from agent $b$ together, or the output from agent $b$ and the readings of sensor $c$ together. 
In addition, each cyber-physical component has an associated attack cost that represents the effort required by an attacker to compromise that component. 
Now, considering these costs, the question we are trying to answer is: which nodes should be compromised in order to disrupt the operation of actuator $c1$, with minimal effort (cost) for the attacker? In other words, what is the least-effort attack strategy to disable actuator $c1$?

Our example involves many attack alternatives, however, only one is minimal. The optimal strategy is to compromise nodes $a$ and $c$ with a total cost of $4$. The compromise of these sensors will disable both AND nodes and consecutively the OR node, which in turn will affect node $d$ and finally node~$c1$. 
\section{MaxSAT formulation strategy}
Given a target node $t$, the input graph $G$ can be used as a map to decode the dependencies that node $t$ relies on. Therefore, $G$ can be traversed backwards in order to produce a propositional formula that represents the different ways in which node $t$ can be fulfilled. We call this transformation $\form(t)$. 
In our example,  $\form(c1)$ is as follows: 
\vspace{-0.05cm}
\begin{lstlisting}[
label={lst:form-step1}, 
mathescape=true, language=bash, %caption=Security metric resolution, 
basicstyle=\martinListingFontsize\sffamily, 
keywordstyle=\bfseries\color{green!40!black}, 
commentstyle=\itshape\color{purple!40!black},
identifierstyle=\color{blue},
stringstyle=\color{orange}, 
frame=single, 
%numbers=left,
numberstyle=\tiny\color{gray},
stepnumber=1, 
belowcaptionskip=5em,
belowskip=3em, 
captionpos=b, 
%xleftmargin=.125\textwidth, xrightmargin=.125\textwidth
]
             $\form(c1) = c1 \land ( d \land ( ( a \land b ) \lor ( b \land c ) ) )$
\end{lstlisting}
\vspace{-0.5cm}

\vspace{-0.2cm}
The goal of the attacker, however, is precisely the opposite, i.e., to disrupt node $c1$ somewhere along the graph. 
Therefore, we are actually interested in satisfying $\neg \form(c1)$, which describes the means to disable $c1$. 
After applying a few logical rules, the conjunctive normal form (CNF) of $\neg \form(c1)$ is: 
\begin{lstlisting}[
label={lst:form-step3}, 
mathescape=true, language=bash, %caption=Security metric resolution, 
basicstyle=\martinListingFontsize\sffamily, 
keywordstyle=\bfseries\color{green!40!black}, 
commentstyle=\itshape\color{purple!40!black},
identifierstyle=\color{blue},
stringstyle=\color{orange}, 
frame=single, 
%numbers=left,
numberstyle=\tiny\color{gray},
stepnumber=1, 
belowcaptionskip=5em,
belowskip=3em, 
captionpos=b, 
%xleftmargin=.125\textwidth, xrightmargin=.125\textwidth
]                   
    $\neg \form(c1) = (\neg c1 \lor \neg d \lor \neg a \lor \neg b) \land (\neg c1 \lor \neg d \lor \neg b \lor \neg c )$
\end{lstlisting}
\vspace{-0.7cm}

In practice, we do not use the naive CNF conversion approach since it might lead to exponential computation times over large graphs.  
Instead, we use the Tseitin transformation~\cite{Tseitin70}, which can be done in polynomial time and essentially produces a new formula in CNF that is not strictly equivalent to the original formula (because there are new variables) but is equisatisfiable. 
This means that, given an assignment of truth values, the new formula is satisfied if and only if the original formula is also satisfied. 
Under that perspective, a logical assignment such that $\neg \form(t) = true$ will indicate which nodes must be compromised (i.e. logically falsified) in order to disrupt the operation of the system. 
 
Considering the CNF formula produced by the Tseitin transformation and a cost function $\cost(n)$ that indicates the attack cost of a node $n$, we model our problem as a Weighted Partial MaxSAT problem \cite{Davies2011}. 
Hard clauses are essentially the clauses within the CNF formula: 

\vspace{0.25cm}
\hspace{0.25cm}
\begin{minipage}{.5\textwidth}
    \centering
    \begin{minipage}{.48\textwidth}
      {\renewcommand{\arraystretch}{1.15}
            \footnotesize{
              \centering
              \begin{tabular}{|c|}
                  \hline
                  $\neg c1 \lor \neg d \lor \neg a \lor \neg b$\\ 
                  \hline
              \end{tabular}
          }
      }
    \end{minipage}%
    \begin{minipage}{0.48\textwidth}
       {\renewcommand{\arraystretch}{1.15}
               \centering
               \footnotesize{
               \begin{tabular}{|c|}
                   \hline
                   $\neg c1 \lor \neg d \lor \neg b \lor \neg c$\\
                   \hline
               \end{tabular}
           }
       }
    \end{minipage}
\vspace{0.2cm}
\end{minipage}

\hspace{-0.34cm}whereas soft clauses correspond to each atomic node in the graph with their corresponding penalties (costs) as follows: 
{\renewcommand{\arraystretch}{1.15}
    \begin{table}[!h]
        \centering
        \begin{tabular}{|c|c|c|c|c|}
            \hline
            {$a$} & {$b$} & {$c$} & {$d$} & {$c1$} \\
            \hline
            {$\cost(a)=2$} & {$\cost(b)=5$} & {$\cost(c)=2$} & {$\cost(d)=10$} & {$\cost(c1)=inf$} \\
            \hline
        \end{tabular}	
    \end{table}
}

Therefore, a MaxSAT solver will try to minimise the number of falsified variables as well as their weights, which in our problem equals to minimise the compromise cost for the attacker. Note: the additional variables introduced by the Tseitin transformation have cost/weight 0 in the formulation. 

\section{AND/OR graph generation} 
The benchmark presented in this paper relies on \toolx, a Java-based security metric analyser for ICS~\cite{Barrere2019}, \cite{BarrereMeta4icsGithub}. We have used \tool to produce and analyse synthetic pseudo-random AND/OR graphs of different size and composition. 
To create an AND/OR graph of size $n$, we first create the target node. Afterwards, we create a predecessor which has one of the three types (atomic, AND, OR) according to a probability given by a compositional configuration predefined for the experiment. 
For example, a configuration of $(60,20,20)$ means 60\% of atomic nodes, 20\% of AND nodes and 20\% of OR nodes. 
We repeat this process creating children on the respective nodes until we approximate the desired size of the graph~$n$. 
Node costs, represented by $\cost(n)$, are integer values randomly selected between 1 and 100. 

The benchmark also includes the solutions obtained by \tool for each case, including resolution time, total cost and critical nodes. Currently, \tool uses SAT4J \cite{SAT4J} and a Python-based linear programming approach as MaxSAT solvers. The tool runs all available solvers in parallel and picks the first one that comes up with a valid solution.

\section{Benchmark description}
Out dataset includes 80 cases in total, and can be obtained at \cite{BarrereMeta4icsGithub}. 
There are four different sizes of AND/OR graphs involving 5000, 10000, 15000, and 20000 nodes (20 cases each). 
For each graph size, we consider two different graph configurations, 80/10/10 and 60/20/20, which determine the composition of the graphs (10 cases each).  
Table \ref{tab:experiments} shows the identifiers of the cases within each one of these categories. 

\begin{table}[h!]
    \centering
    \begin{tabular}{@{}|c|c|c|@{}}
        \toprule
        \textbf{Nodes/Configurations} & \textbf{80/10/10 config} & \textbf{60/20/20 config} \\ \midrule
        5000 & 1 to 10 & 11 to 20 \\ \midrule
        10000 & 21 to 30 & 31 to 40 \\ \midrule
        15000 & 41 to 50 & 51 to 60 \\ \midrule
        20000 & 61 to 70 & 71 to 80 \\ \bottomrule
    \end{tabular}
    \vspace{0.5cm}
    \caption{Benchmark cases and configurations}
    \label{tab:experiments}
\end{table}
\vspace{-0.5cm}

Each case is specified in an individual \textbf{.wcnf} (DIMACS-like, weighted CNF) file named with the case id and the number of nodes involved. 
The weight for hard clauses (\textit{top} value) has been set to $1.0 \times 10^ 6$. 
Tables \ref{tab:benchmark-part1} and \ref{tab:benchmark-part2} detail each case as well as the results obtained with our tool. 
The field \textbf{id} identifies each case; 
\textbf{gNodes} indicates the total number of nodes in the original AND/OR graph; 
\textbf{gAT}, \textbf{gAND} and \textbf{gOR} indicate the approximate composition of the graph in terms of atomic (cyber-physical components), AND and OR nodes; 
\textbf{tsVars} and \textbf{tsClauses} show the number of variables and clauses involved in the MaxSAT formulation after applying the Tseitin transformation;  
\textbf{cost} and \textbf{time} show the total solution cost reported by \tool and the time needed for its resolution in milliseconds; 
\textbf{solution} shows the set of critical nodes expressed as a list of nodes with their respective costs~(weights). 
These experiments have performed on a MacBook Pro (15-inch, 2018), 2.9 GHz Intel Core i9, 32 GB 2400 MHz DDR4.

As a final remark, it can be observed that some cases with the same size and composition parameters have very different resolution times. 
This is an interesting phenomenon and it is due to the internal logical composition of the AND/OR graph and how well the underlying solver performs with each case. Within our experiments, we have observed that none of the two solvers used in \tool is faster than the other in all of the cases. We believe this is an interesting problem that should be further investigated in the context of MaxSAT solvers. 
\begin{table*}[]
	\centering
	\begin{tabular}{@{}|c|c|c|c|c|c|c|c|c|c|@{}}
		\toprule
		\textbf{id} & \textbf{gNodes} & \textbf{gAT} & \textbf{gAND} & \textbf{gOR} & \textbf{tsVars} & \textbf{tsClauses} & \textbf{cost} & \textbf{time} & \textbf{solution}                                                                                                                                                                                                                                                                           \\ \midrule
		1           & 5000            & 3986         & 540           & 475          & 8987            & 23990              & 20            & 1045          & {[}(2:9),(5120:4),(8900:4),(8904:3){]}                                                                                                                                                                                                                                                 \\ \midrule
		2           & 5000            & 4015         & 515           & 471          & 9016            & 24019              & 3             & 863           & {[}(940:1),(3390:1),(3424:1){]}                                                                                                                                                                                                                                                        \\ \midrule
		3           & 5000            & 3997         & 514           & 490          & 8998            & 24001              & 1             & 847           & {[}(7950:1){]}                                                                                                                                                                                                                                                                         \\ \midrule
		4           & 5000            & 3982         & 506           & 513          & 8983            & 23986              & 3             & 850           & {[}(874:2),(5081:1){]}                                                                                                                                                                                                                                                                 \\ \midrule
		5           & 5000            & 4018         & 503           & 480          & 9019            & 24022              & 2             & 837           & {[}(6:2){]}                                                                                                                                                                                                                                                                            \\ \midrule
		6           & 5000            & 4013         & 514           & 474          & 9014            & 24017              & 3             & 855           & {[}(5235:3){]}                                                                                                                                                                                                                                                                         \\ \midrule
		7           & 5000            & 4004         & 500           & 497          & 9005            & 24008              & 1             & 840           & {[}(4:1){]}                                                                                                                                                                                                                                                                            \\ \midrule
		8           & 5000            & 3999         & 508           & 494          & 9000            & 24003              & 5             & 831           & {[}(8796:5){]}                                                                                                                                                                                                                                                                         \\ \midrule
		9           & 5000            & 4010         & 502           & 489          & 9011            & 24014              & 2             & 827           & {[}(6219:2){]}                                                                                                                                                                                                                                                                         \\ \midrule
		10          & 5000            & 3973         & 531           & 497          & 8974            & 23977              & 9             & 838           & {[}(7:3),(6319:1),(6859:5){]}                                                                                                                                                                                                                                                          \\ \midrule
		11          & 5000            & 3008         & 1025          & 968          & 8009            & 23012              & 8             & 825           & {[}(1033:6),(4416:2){]}                                                                                                                                                                                                                                                                \\ \midrule
		12          & 5000            & 3020         & 1020          & 961          & 8021            & 23024              & 1             & 820           & {[}(3366:1){]}                                                                                                                                                                                                                                                                         \\ \midrule
		13          & 5000            & 3017         & 969           & 1015         & 8018            & 23021              & 100           & 2157          & {[}(2:6),(5192:7),(6527:28),(7195:3),(7236:20),(7500:7),\\
		&             &          &           &           &             &               &            &          &(7556:10),(7919:10),(7943:3),(7950:2),(7972:4){]}                                                                                                                                                                              \\ \midrule
		14          & 5000            & 3026         & 981           & 994          & 8027            & 23030              & 3             & 823           & {[}(2:3){]}                                                                                                                                                                                                                                                                            \\ \midrule
		15          & 5000            & 3027         & 971           & 1003         & 8028            & 23031              & 16            & 823           & {[}(2:16){]}                                                                                                                                                                                                                                                                           \\ \midrule
		16          & 5000            & 3004         & 1013          & 984          & 8005            & 23008              & 2             & 826           & {[}(7497:2){]}                                                                                                                                                                                                                                                                         \\ \midrule
		17          & 5000            & 3017         & 1053          & 931          & 8018            & 23021              & 7             & 827           & {[}(5844:7){]}                                                                                                                                                                                                                                                                         \\ \midrule
		18          & 5000            & 3021         & 996           & 984          & 8022            & 23025              & 22            & 822           & {[}(6103:5),(6119:17){]}                                                                                                                                                                                                                                                               \\ \midrule
		19          & 5000            & 3022         & 991           & 988          & 8023            & 23026              & 26            & 828           & {[}(4118:7),(4144:19){]}                                                                                                                                                                                                                                                               \\ \midrule
		20          & 5000            & 3011         & 988           & 1002         & 8012            & 23015              & 17            & 816           & {[}(17:10),(549:4),(5846:3){]}                                                                                                                                                                                                                                                         \\ \midrule
		21          & 10000           & 8017         & 1021          & 963          & 18018           & 48021              & 1             & 1033          & {[}(456:1){]}                                                                                                                                                                                                                                                                          \\ \midrule
		22          & 10000           & 7991         & 1003          & 1007         & 17992           & 47995              & 17            & 911           & {[}(2:17){]}                                                                                                                                                                                                                                                                           \\ \midrule
		23          & 10000           & 8023         & 987           & 991          & 18024           & 48027              & 2             & 894           & {[}(365:2){]}                                                                                                                                                                                                                                                                          \\ \midrule
		24          & 10000           & 8035         & 967           & 999          & 18036           & 48039              & 57            & 4398          & {[}(10137:2),(11192:2),(14319:1),(4:14),(607:2),(7746:9),\\
		&             &          &           &           &             &               &            &          & (7780:2),(7871:2),(8108:12),(8307:5),(9879:2),(9909:4){]}                                                                                                                                                                     \\ \midrule
		25          & 10000           & 7996         & 1054          & 951          & 17997           & 48000              & 4             & 927           & {[}(6:4){]}                                                                                                                                                                                                                                                                            \\ \midrule
		26          & 10000           & 7967         & 1006          & 1028         & 17968           & 47971              & 9             & 908           & {[}(11920:7),(17849:2){]}                                                                                                                                                                                                                                                              \\ \midrule
		27          & 10000           & 7954         & 1038          & 1009         & 17955           & 47958              & 2             & 858           & {[}(14335:2){]}                                                                                                                                                                                                                                                                        \\ \midrule
		28          & 10000           & 8013         & 1011          & 977          & 18014           & 48017              & 22            & 917           & {[}(9:6),(428:1),(549:4),(679:7),(8781:2),(15313:2){]}                                                                                                                                                                                                                                 \\ \midrule
		29          & 10000           & 8008         & 967           & 1026         & 18009           & 48012              & 5             & 862           & {[}(149:3),(190:2){]}                                                                                                                                                                                                                                                                  \\ \midrule
		30          & 10000           & 8026         & 1023          & 952          & 18027           & 48030              & 1             & 884           & {[}(4553:1){]}                                                                                                                                                                                                                                                                         \\ \midrule
		31          & 10000           & 6035         & 1996          & 1970         & 16036           & 46039              & 289           & 4206          & {[}(13217:8),(15933:12),(2:29),(2362:2),(2573:1),(3452:20),\\
		&             &          &           &           &             &               &            &          & (3510:8),(3552:35),(4168:12),(4187:29),(4226:1),(4438:78),\\
		&             &          &           &           &             &               &            &          &(4664:51),(5550:2),(6920:1){]}                                                                                                                                    \\ \midrule
		32          & 10000           & 6028         & 1989          & 1984         & 16029           & 46032              & 22            & 859           & {[}(3:22){]}                                                                                                                                                                                                                                                                           \\ \midrule
		33          & 10000           & 6058         & 2015          & 1928         & 16059           & 46062              & 1             & 849           & {[}(13604:1){]}                                                                                                                                                                                                                                                                        \\ \midrule
		34          & 10000           & 6012         & 2001          & 1988         & 16013           & 46016              & 3             & 850           & {[}(11999:3){]}                                                                                                                                                                                                                                                                        \\ \midrule
		35          & 10000           & 6055         & 1958          & 1988         & 16056           & 46059              & 8             & 851           & {[}(2:8){]}                                                                                                                                                                                                                                                                            \\ \midrule
		36          & 10000           & 5993         & 1954          & 2054         & 15994           & 45997              & 33            & 855           & {[}(14291:2),(14602:28),(14991:3){]}                                                                                                                                                                                                                                                   \\ \midrule
		37          & 10000           & 6064         & 1927          & 2010         & 16065           & 46068              & 4             & 850           & {[}(6:4){]}                                                                                                                                                                                                                                                                            \\ \midrule
		38          & 10000           & 6015         & 1993          & 1993         & 16016           & 46019              & 289           & 4965          & {[}(10212:1),(10217:41),(10224:14),(10253:21),(10303:4),(10588:8),\\
		&             &          &           &           &             &               &            &          &(11576:19),(11974:36),(12267:10),(12396:3),(12724:1),(12794:21),\\
		&             &          &           &           &             &               &            &          & (12846:4),(12934:15),(13374:1),(13850:17),(14318:19),(15239:1),\\
		&             &          &           &           &             &               &            &          & (15272:2),(3169:2),(3794:20),(3862:1),(3934:8),(4090:8),\\
		&             &          &           &           &             &               &            &          &(8243:5),(8248:4),(9905:3){]} \\ \midrule
		39          & 10000           & 6012         & 1946          & 2043         & 16013           & 46016              & 4             & 856           & {[}(11750:4){]}                                                                                                                                                                                                                                                                        \\ \midrule
		40          & 10000           & 6042         & 1956          & 2003         & 16043           & 46046              & 1             & 858           & {[}(4548:1){]}                                                                                                                                                                                                                                                                         \\ \bottomrule
	\end{tabular}
	 \vspace{0.5cm}
	\caption{Benchmark description - Cases 1 to 40}
	\label{tab:benchmark-part1}
\end{table*}

\begin{table*}[]
	\centering
	\begin{tabular}{@{}|c|c|c|c|c|c|c|c|c|c|@{}}
		\toprule
		\textbf{id} & \textbf{gNodes} & \textbf{gAT} & \textbf{gAND} & \textbf{gOR} & \textbf{tsVars} & \textbf{tsClauses} & \textbf{cost} & \textbf{time} & \textbf{solution}                                                                                                                                                                                             \\ \midrule
		41          & 15000           & 11999        & 1498          & 1504         & 27000           & 72003              & 5             & 1152          & {[}(3095:2),(25236:1),(25957:2){]}                                                                                                                                                                       \\ \midrule
		42          & 15000           & 12016        & 1527          & 1458         & 27017           & 72020              & 2             & 982           & {[}(26790:2){]}                                                                                                                                                                                          \\ \midrule
		43          & 15000           & 11983        & 1454          & 1564         & 26984           & 71987              & 6             & 1095          & {[}(23:4),(25147:2){]}                                                                                                                                                                                   \\ \midrule
		44          & 15000           & 11936        & 1561          & 1504         & 26937           & 71940              & 1             & 939           & {[}(6:1){]}                                                                                                                                                                                              \\ \midrule
		45          & 15000           & 12006        & 1528          & 1467         & 27007           & 72010              & 1             & 955           & {[}(4963:1){]}                                                                                                                                                                                           \\ \midrule
		46          & 15000           & 11995        & 1465          & 1541         & 26996           & 71999              & 34            & 2968          & {[}(6:25),(22122:1),(22891:6),(23040:2){]}                                                                                                                                                               \\ \midrule
		47          & 15000           & 11973        & 1507          & 1521         & 26974           & 71977              & 2             & 932           & {[}(4:2){]}                                                                                                                                                                                              \\ \midrule
		48          & 15000           & 12021        & 1490          & 1490         & 27022           & 72025              & 36            & 1092          & {[}(76:4),(416:20),(2148:1),(9656:4),(17438:1),(23881:6){]}                                                                                                                                              \\ \midrule
		49          & 15000           & 11957        & 1528          & 1516         & 26958           & 71961              & 4             & 916           & {[}(14464:2),(24824:2){]}                                                                                                                                                                                \\ \midrule
		50          & 15000           & 12009        & 1518          & 1474         & 27010           & 72013              & 10            & 992           & {[}(10163:1),(19872:1),(22070:3),(26038:5){]}                                                                                                                                                            \\ \midrule
		51          & 15000           & 9057         & 2979          & 2965         & 24058           & 69061              & 214           & 6335          & {[}(11377:5),(16191:11),(17087:52),(17422:3),(18195:16),(18367:4),\\
				&             &          &           &           &             &               &            &          & (18383:1),(18593:6),(18991:3),(19054:5),(19448:3),(19551:32),\\
				&             &          &           &           &             &               &            &          & (21785:1),(21807:10),(22034:8),(22087:32),(23437:15),\\
				&             &          &           &           &             &               &            &          &  (6815:3),(6919:4){]} \\ \midrule
		52          & 15000           & 9072         & 2949          & 2980         & 24073           & 69076              & 49            & 1024          & {[}(2:1),(14480:7),(17852:41){]}                                                                                                                                                                         \\ \midrule
		53          & 15000           & 9023         & 3072          & 2906         & 24024           & 69027              & 10            & 940           & {[}(1425:6),(20681:4){]}                                                                                                                                                                                 \\ \midrule
		54          & 15000           & 9062         & 2973          & 2966         & 24063           & 69066              & 45            & 2000          & {[}(6912:4),(9199:25),(12387:13),(13814:2),(14190:1){]}                                                                                                                                                  \\ \midrule
		55          & 15000           & 9006         & 2983          & 3012         & 24007           & 69010              & 100           & 4397          & {[}(1496:1),(4027:6),(5220:2),(5781:1),(16796:89),(21242:1){]}                                                                                                                                           \\ \midrule
		56          & 15000           & 9050         & 2936          & 3015         & 24051           & 69054              & 138           & 7839          & {[}(11358:4),(1460:7),(16765:3),(3845:81),(666:8),\\
		&             &          &           &           &             &               &            &          &  (7936:1),(8733:8),(8870:18),(8959:8){]}                                                                                                                \\ \midrule
		57          & 15000           & 9039         & 2996          & 2966         & 24040           & 69043              & 11            & 924           & {[}(12114:9),(12732:2){]}                                                                                                                                                                                \\ \midrule
		58          & 15000           & 9057         & 2967          & 2977         & 24058           & 69061              & 66            & 8208          & {[}(16552:13),(2:53){]}                                                                                                                                                                                  \\ \midrule
		59          & 15000           & 9047         & 3034          & 2920         & 24048           & 69051              & 11            & 1015          & {[}(3226:4),(21769:7){]}                                                                                                                                                                                 \\ \midrule
		60          & 15000           & 9036         & 2963          & 3002         & 24037           & 69040              & 116           & 8376          & {[}(11185:6),(12042:16),(12906:26),(13556:1),\\
		&             &          &           &           &             &               &            &          & (13652:7),(14234:8),(2:52){]}                                                                                                                               \\ \midrule
		61          & 20000           & 16012        & 2010          & 1979         & 36013           & 96016              & 16            & 4456          & {[}(77:2),(7769:5),(18831:1),(24368:8){]}                                                                                                                                                                \\ \midrule
		62          & 20000           & 15998        & 2039          & 1964         & 35999           & 96002              & 1             & 1020          & {[}(12:1){]}                                                                                                                                                                                             \\ \midrule
		63          & 20000           & 15962        & 2045          & 1994         & 35963           & 95966              & 14            & 1134          & {[}(1370:6),(2437:1),(9642:1),(26442:4),(28443:2){]}                                                                                                                                                     \\ \midrule
		64          & 20000           & 15961        & 2026          & 2014         & 35962           & 95965              & 4             & 1004          & {[}(14:4){]}                                                                                                                                                                                             \\ \midrule
		65          & 20000           & 16053        & 1999          & 1949         & 36054           & 96057              & 2             & 1014          & {[}(11:2){]}                                                                                                                                                                                             \\ \midrule
		66          & 20000           & 16028        & 1983          & 1990         & 36029           & 96032              & 1             & 960           & {[}(30735:1){]}                                                                                                                                                                                          \\ \midrule
		67          & 20000           & 15983        & 2050          & 1968         & 35984           & 95987              & 1             & 1020          & {[}(8:1){]}                                                                                                                                                                                              \\ \midrule
		68          & 20000           & 15979        & 2006          & 2016         & 35980           & 95983              & 20            & 4334          & {[}(1147:2),(3275:2),(7063:1),(7351:5),(9616:3),\\
		&             &          &           &           &             &               &            &          &  (14569:1),(23153:4),(25514:1),(26598:1){]}                                                                                                               \\ \midrule
		69          & 20000           & 16008        & 1990          & 2003         & 36009           & 96012              & 18            & 2736          & {[}(5:18){]}                                                                                                                                                                                             \\ \midrule
		70          & 20000           & 16119        & 1921          & 1961         & 36120           & 96123              & 5             & 978           & {[}(4:5){]}                                                                                                                                                                                              \\ \midrule
		71          & 20000           & 12068        & 3848          & 4085         & 32069           & 92072              & 25            & 1034          & {[}(2:25){]}                                                                                                                                                                                             \\ \midrule
		72          & 20000           & 12028        & 3962          & 4011         & 32029           & 92032              & 3             & 983           & {[}(31797:3){]}                                                                                                                                                                                          \\ \midrule
		73          & 20000           & 12054        & 4046          & 3901         & 32055           & 92058              & 7             & 1006          & {[}(22596:5),(22673:1),(24195:1){]}                                                                                                                                                                      \\ \midrule
		74          & 20000           & 12051        & 4033          & 3917         & 32052           & 92055              & 28            & 1094          & {[}(26825:4),(29603:3),(30184:20),(30210:1){]}                                                                                                                                                           \\ \midrule
		75          & 20000           & 12103        & 3993          & 3905         & 32104           & 92107              & 5             & 965           & {[}(19799:5){]}                                                                                                                                                                                          \\ \midrule
		76          & 20000           & 12000        & 4004          & 3997         & 32001           & 92004              & 9             & 995           & {[}(15575:3),(17158:6){]}                                                                                                                                                                                \\ \midrule
		77          & 20000           & 12127        & 3851          & 4023         & 32128           & 92131              & 18            & 976           & {[}(2821:3),(4451:13),(9408:2){]}                                                                                                                                                                        \\ \midrule
		78          & 20000           & 11957        & 4006          & 4038         & 31958           & 91961              & 2             & 938           & {[}(12459:1),(20894:1){]}                                                                                                                                                                                \\ \midrule
		79          & 20000           & 12073        & 3964          & 3964         & 32074           & 92077              & 49            & 1757          & {[}(20965:47),(24362:2){]}                                                                                                                                                                               \\ \midrule
		80          & 20000           & 12048        & 3940          & 4013         & 32049           & 92052              & 50            & 1432          & {[}(2537:4),(7471:2),(7483:8),(15594:5),(22496:1),(23848:2),\\
		&             &          &           &           &             &               &            &          & (27795:9),(28554:19){]}                                                                                                                      \\ \bottomrule
	\end{tabular}
	\vspace{0.5cm}
	\caption{Benchmark description - Cases 41 to 80}
	\label{tab:benchmark-part2}
\end{table*}

\bibliographystyle{IEEEtran}
\bibliography{IEEEabrv,doc.bib}

\end{document}